\documentclass[preprint,12p]{revtex4}

\usepackage[dvipsnames]{xcolor}
\usepackage[brazil, english]{babel}
\usepackage[utf8]{inputenc}
\usepackage{graphicx}
\usepackage{epsfig}
\usepackage{amssymb}
\usepackage{amsmath} 
\usepackage{soul}
\usepackage[unicode=true,bookmarks=false,breaklinks=false,pdfborder={0 0 1},colorlinks=true]
 {hyperref}
\hypersetup{
 citecolor=blue,linkcolor=blue,urlcolor=blue}

\graphicspath{%
    {converted_graphics/}
    {/}
}

\begin{document}

\title{Joule-Thomson expansion for noncommutative\\ uncharged black holes}
\author{J. P. Morais Graça$^{1}$}
\email[Eletronic address: ]{jpmorais@gmail.com}
\author{Eduardo Folco Capossoli$^{2}$}
\email[Eletronic address: ]{eduardo\_capossoli@cp2.g12.br}
\author{Henrique Boschi-Filho$^{1}$}
\email[Eletronic address: ]{boschi@if.ufrj.br}  
\affiliation{$^1$Instituto de F\'{\i}sica, Universidade Federal do Rio de Janeiro, 21.941-972 - Rio de Janeiro-RJ - Brazil \\ 
 $^2$Departamento de F\'{\i}sica and Mestrado Profissional em Práticas de Educa\c{c}\~{a}o B\'{a}sica (MPPEB),  Col\'egio Pedro II, 20.921-903 - Rio de Janeiro-RJ - Brazil
 }

\begin{abstract}
 In this work we study the Joule-Thomson expansion for uncharged black holes in a noncommutative scenario characterized by a parameter $\theta$, which is present in the horizon function. We calculate the inversion temperature for some values of $\theta$ and the isenthalpics for fixed masses. We find that the uncharged noncommutative black hole  behaves as a charged commutative one. 
\end{abstract}

\maketitle


\section{Introduction}

 Classical black holes, as solutions of Einstein equations, are characterized by just a few quantities (mass, charge, angular momentum).  
 When quantum processes are incorporated into their  study, one finds that such objects behave as  thermodynamic systems and radiate like a black body \cite{Hawking:1974sw}.
To compensate the energy that is lost in this process, its temperature increases with the decrease of its radius, in such a way as to lead to a catastrophic behavior in its final stage, the evaporation. 
However, at this stage it is expected that processes linked to quantum gravity must be considered. 

In the context of string theory, more precisely within the black holes/string correspondence \cite{Susskind:1993ki}, it has been proposed that we can take into account the quantization of spacetime through the idea that the coordinates in the target space become noncommutative \cite{Witten:1995im,Seiberg:1999vs}. Based on this idea, Nicolini, Smailagic and Spallucci studied the evaporation of a Schwarzschild-like black hole when spacetime is noncommutative \cite{Nicolini:2005vd} and found that, rather than a catastrophic behavior, the final stage of such a black hole is actually a finite-mass cold remnant.

The noncommutativity of spacetime can be described via the commutator 
\begin{equation}
    [x^\mu, x^\nu] = \theta\,  \epsilon^{\mu\nu}\,, 
\end{equation}

\noindent
where $\epsilon^{\mu\nu}$ is an antisymmetric matrix and $\theta$ is the parameter that will measure the noncommutativity of the coordinates. Nicolini et al \cite{Nicolini:2005vd} proposed that this noncommutativity can be incorporated into gravitation without changing Einstein's equations, but modifying the energy-momentum tensor to include such effects directly in the source of the curvature. In this way, a collapsed object like a black hole would not be described by a singularity, but by a kind of anisotropic fluid, diffused over a certain space region proportional to the parameter $\theta$. 

Originally, this distribution was treated in a Gaussian way, for simplicity's sake, but non-Gaussian black holes, based on the noncommutativity of spacetime, were proposed afterwards (see, for instance, \cite{Park:2008ud}). The thermodynamics of such objects has already been studied \cite{Banerjee:2008gc,Miao:2016ulg}, as well as other properties \cite{Ding:2010dc,Wei:2015dua,Gupta:2015uga,Brown:2010cr,Anacleto:2014cga, Anacleto:2014apa,Gupta:2017lwk,Miao:2011vdq,Anacleto:2019tdj}. In this paper, we will focus on the Joule-Thomson expansion of noncommutative black holes, based on what is called black hole chemistry, where the mass of the black hole is identified with its enthalpy \cite{Kastor:2009wy}. 

Black hole chemistry is based on the idea that black holes have a mechanical pressure and a thermodynamic volume among their state variables (for a review on the subject, see \cite{Kubiznak:2016qmn}). The main reason for this proposal is perhaps based on the fact that, considering the cosmological constant as proportional to the pressure, its conjugate variable is exactly the volume of a sphere whose radius is the horizon radius \cite{Kastor:2009wy, Kubiznak:2012wp} for the Schwarzschild AdS metric. 

Among the processes studied in the context of black hole chemistry, the Joule-Thomson expansion analyzes the black hole temperature variation by its pressure, keeping the enthalpy of the system constant \cite{Okcu:2016tgt,Okcu:2017qgo,Mo:2018rgq,Lan:2018nnp,Mo:2018qkt,Cisterna:2018jqg,Rizwan:2018mpy,Chabab:2018zix, Ghaffarnejad:2018exz, Yekta:2019wmt,Li:2019jcd,Nam:2019zyk, K.:2020rzl, Bi:2020vcg, Graca:2021izb, Yin:2021akt, Biswas:2021uop}. Thus, it is worth studying different metrics for black holes  to look for some similarity with fluid/gas systems, as van der Waals ones.   

An interesting result is that Schwarzschild AdS black holes do not have temperature inversion curves unless some kind of electrical charge is included in the system, that is, for a charged black hole like the Reissner-Nordström type. This brings us to the question of whether there are black holes that have temperature inversion curves even when not charged. The main motivation of this work is to investigate whether uncharged noncommutative black holes satisfy this criterion, i.e., that they have or not these inversion temperatures. The result obtained here is that noncommutative black holes do have these inversion temperature curves, and we also present their behaviors throughout $T-P$ phase diagrams.

The paper is arranged as follows. In section \ref{JTBHNC} we establish the noncommutative black hole scenario, describe the Joule-Thomson expansion which is related to the inversion temperature. In section \ref{num}  we present and discuss our numerical results for the inversion temperature curves as well as the isenthalpics in $T-P$ phase diagrams. In section \ref{conc} we bring our conclusion and last comments. 

\section{Joule-Thomson expansion in non-commutative black holes}\label{JTBHNC}

Let us start this section introducing the metric for noncommutative black hole as proposed by Nicolini et al \cite{Nicolini:2005vd} as
\begin{equation}
ds^2 = - f(r)\,  dt^2 + \frac{ dr^2}{f(r)}+ r^2\, d\Omega^2,
\end{equation}

\noindent
where $d\Omega^2$ is the solid angle,  $f(r)$, the so-called horizon function, 
\begin{equation}
    f(r)=1 -\frac{4 M }{\sqrt{\pi }\, r}\,\gamma \left(\frac{3}{2},\frac{r^2}{4 \theta }\right) + \frac{r^2}{L^2},
    \label{funcaocompleta}
\end{equation}

\noindent
and $\gamma(a,x)$ is the lower incomplete gamma function, defined by
\begin{equation}
    \gamma (a, x) = \int_0^{x} dt \, t^{a-1} \, e^{-t}\,. 
\end{equation}
As stated in the introduction, the parameter $\theta$ measures the noncommutativity of spacetime. The parameter $M$ is the mass of the black hole, and the other parameter of the above metric, $L$, is the radius of the anti-de Sitter spacetime. 

To get a feeling of the role of the noncommutative paremeter $\theta$ in the black hole metric let us expand it for small $\theta$, as
\begin{eqnarray}
 f(r) &\approx& 1 -\frac{2 M}{r}+\frac{2 M e^{-\frac{r^2}{4 \theta }}}{\sqrt{\pi } \sqrt{\theta }}+\frac{4 \sqrt{\theta } M e^{-\frac{r^2}{4 \theta }}}{\sqrt{\pi } r^2} + \frac{r^2}{L^2}.
 \label{funcaoaproximada}
\end{eqnarray}
and for large $r^2/4\theta$, one has a Taylor expansion for $e^{-x^2}$ around $x_0=M/\sqrt{\theta}$ (large) which means $r\approx 2M$ (classical Schwarzschild radius): 
\begin{eqnarray}
 e^{- \frac{r^2}{4\theta}}
 = e^{- \frac{M^2}{\theta}}\left[ 
 1 - \frac {Mr}{\theta} + \frac {2M^2}{\theta} + \dots 
 \right]
\end{eqnarray}
so that
\begin{eqnarray}
 f(r) &\approx& 1 -\frac{2 M}{r}+
 \left[ \frac{2 M }{\sqrt{\pi } \sqrt{\theta }}+\frac{4 \sqrt{\theta } M }{\sqrt{\pi } r^2} \right]e^{- \frac{M^2}{\theta}}\left[ 
 1 - \frac {Mr}{\theta} + \frac {2M^2}{\theta} + \dots 
 \right]+ \frac{r^2}{L^2}.
\end{eqnarray}

\noindent
As one can see, in the above expression there is a $1/r^2$ term which resembles a charge-like term, but whose charge would be dependent on the black hole mass. This is an indication that the noncommutative AdS black hole share some properties with a Reissner-Nordström AdS black hole. From now on, we will only work with the exact expression for the event horizon function, equation (\ref{funcaocompleta}).

Following \cite{Kastor:2009wy}, the mass $M$ of the black hole is identified with its enthalpy, then the first law of thermodynamics can be written as
\begin{equation}
    \label{firstLaw}
    dM\equiv dH = T dS + V dP\,
\end{equation}
and the thermodynamic volume can be calculated by 
\begin{equation}\label{vol}
    V = \left( \frac{\partial M}{\partial P} \right)_{S}\,. 
\end{equation}
The mechanical pressure is defined as
\begin{equation}
    P = \frac{3}{8 \pi} \frac{1}{L^2}\,, 
\end{equation}
so that for a Reissner-Nordström AdS black hole the volume would be the canonical one, $V = (4/3) \pi r_+^3$, where $r_+$ is the horizon radius. Note, however, that the volume of the noncommutative black hole, given by \eqref{vol}, will differ from the canonical one. To see this, let us calculate  
the black hole mass $M$, from $f(r_+)=0$, 
\begin{equation}
    M( P, r_+)=\frac{2 \pi ^{3/2} P \, r_+ }{3 \gamma \left(\frac{3}{2},\frac{r_+^2}{4 \theta}\right)}\,
    \left(\frac{3}{8 \pi  P}+r_+^2\right),
\end{equation}

\noindent
so that the thermodynamic volume of the system is given by: 
\begin{equation}
   V = \left( \frac{\partial M}{\partial P} \right)_{S} =  \frac{2 \, r_+^3 \, \pi^{3/2}}{3 \, \gamma\left(\frac{3}{2},\frac{r^2}{4 \theta}\right)}.
\end{equation}

Our interest, in this paper, is to study the  inversion temperature in the Joule-Thomson (JT) expansion. To do this, we start defining the JT coefficient, given by
\begin{equation}
    \mu = \left(\frac{\partial T}{\partial P} \right)_M = \frac{1}{C_P} \left[T \left( \frac{\partial V}{\partial T} \right)_P - V \right],
    \label{JouleThomson}
\end{equation}
\noindent
which tells us how the temperature varies with increasing (or decreasing) pressure. The inversion temperature occurs when $\mu = 0$, i.e., when the system stops heating and starts to cool (or vice versa).

Banerjee, Majhi and Samanta, studying the thermodynamics of noncommutative black holes, concluded that the Bekenstein expression is valid in this context, i.e., $S = A/4$ \cite{Banerjee:2008gc}, so we can calculate the Hawking temperature as
\begin{equation}
    T_H=\frac{\partial M}{\partial S},
\end{equation}

\noindent
which gives us
\begin{equation}\label{thawk}
 T_H=   \frac{{\theta ^{3/2}} 
 \left( \pi  P \, r_+^2 +\frac{1}{8} \right) 
 \gamma \left(\frac{3}{2},\frac{r_+^2}{4 \theta }\right)
 -{\frac{1}{12} r_+^3 \left( \pi  P\, r_+^2 + \frac{3}{8}\right) e^{-\frac{r_+^2}{4 \theta }}}}{ \sqrt{\pi }\,\, {\theta ^{3/2}}\, r_+ \, \gamma \left(\frac{3}{2},\frac{r_+^2}{4 \theta }\right)^2}.
\end{equation}
Using the condition $\mu=0$ in Eq.  (\ref{JouleThomson}), we obtain the inversion temperature as
\begin{eqnarray}\label{tinv}
\nonumber
T_i &=& V\left(\frac{\partial T}{\partial V}\right)_P
\\
&=& \frac{e^{-\frac{r_+^2}{4 \theta }} \left(24 \theta ^3 \left(8 \pi  P r_+^2-1\right) e^{\frac{r_+^2}{2 \theta }} \gamma \left(\frac{3}{2},\frac{r_+^2}{4 \theta }\right)^2+r_+^6 \left(8 \pi  P r_+^2+3\right)\right)}{48 \sqrt{\pi } \theta ^{3/2} \gamma \left(\frac{3}{2},\frac{r_+^2}{4 \theta }\right)^2 \left(12 \theta ^{3/2} r_+ e^{\frac{r_+^2}{4 \theta }} \gamma \left(\frac{3}{2},\frac{r_+^2}{4 \theta }\right)-r_+^4\right)}
\nonumber \\
&+&\frac{e^{-\frac{r_+^2}{4 \theta }} \left(+\sqrt{\theta } r_+^3 e^{\frac{r_+^2}{4 \theta }} \left(-18 \theta +8 \pi  P r_+^4+r_+^2 (3-112 \pi  \theta  P)\right) \gamma \left(\frac{3}{2},\frac{r_+^2}{4 \theta }\right)\right)}{48 \sqrt{\pi } \theta ^{3/2} \gamma \left(\frac{3}{2},\frac{r_+^2}{4 \theta }\right)^2 \left(12 \theta ^{3/2} r_+ e^{\frac{r_+^2}{4 \theta }} \gamma \left(\frac{3}{2},\frac{r_+^2}{4 \theta }\right)-r_+^4\right)}
\end{eqnarray}

Imposing that the inversion temperature of the above equation and the Hawking temperature, Eq.  (\ref{thawk}), are the same when $P = P_i$, we can calculate their difference as: 
\begin{eqnarray}
\Delta T =  \frac {1}{12\sqrt{\pi }} \Bigg\{ &-& 24\, \theta ^{3/2} \left(4 \pi  P r_+^2+1\right) e^{\frac{r_+^2}{4 \theta }} 
+ \, r_+^6 \theta ^{-3/2} \left( \pi  P r_+^2+ \frac{3}{8}\right) e^{-\frac{r_+^2}{4 \theta }} \, \gamma \left(\frac{3}{2},\frac{r_+^2}{4 \theta }\right)^{-2} \cr
&+& 
\frac{1}{2\,\theta}  r_+^3
\left[\frac{3}{4} \,\theta + \pi  P r_+^4+ r_+^2 (\frac{3}{8}- 2 \pi  \theta  P)\right] \gamma \left(\frac{3}{2},\frac{\text{rh}^2}{4 \theta }\right)^{-1} \Bigg\} \cr\cr
&\times &\left[{ \, 
 12\, \theta ^{3/2} r_+  \gamma \left(\frac{3}{2},\frac{r_+^2}{4 \theta }\right)-r_+^4}\right]^{-1} = 0\,,
\end{eqnarray}
which is an equation for $r_+$ and $P$. Solving it  numerically for $r_+$, we can obtain the inversion temperature using Eq. (\ref{thawk}). These results will be presented and discussed in the next section. 

\section{Numerical analysis}\label{num}

 In this section we present the numerical solutions for the inversion temperature as a function of the pressure for some values of $\theta$. Besides, we also present insenthalpic curves ($M$= constant) in $T-P$ diagrams. 
 
\begin{figure}[!ht]
  \centering
  \includegraphics[scale = 0.29]{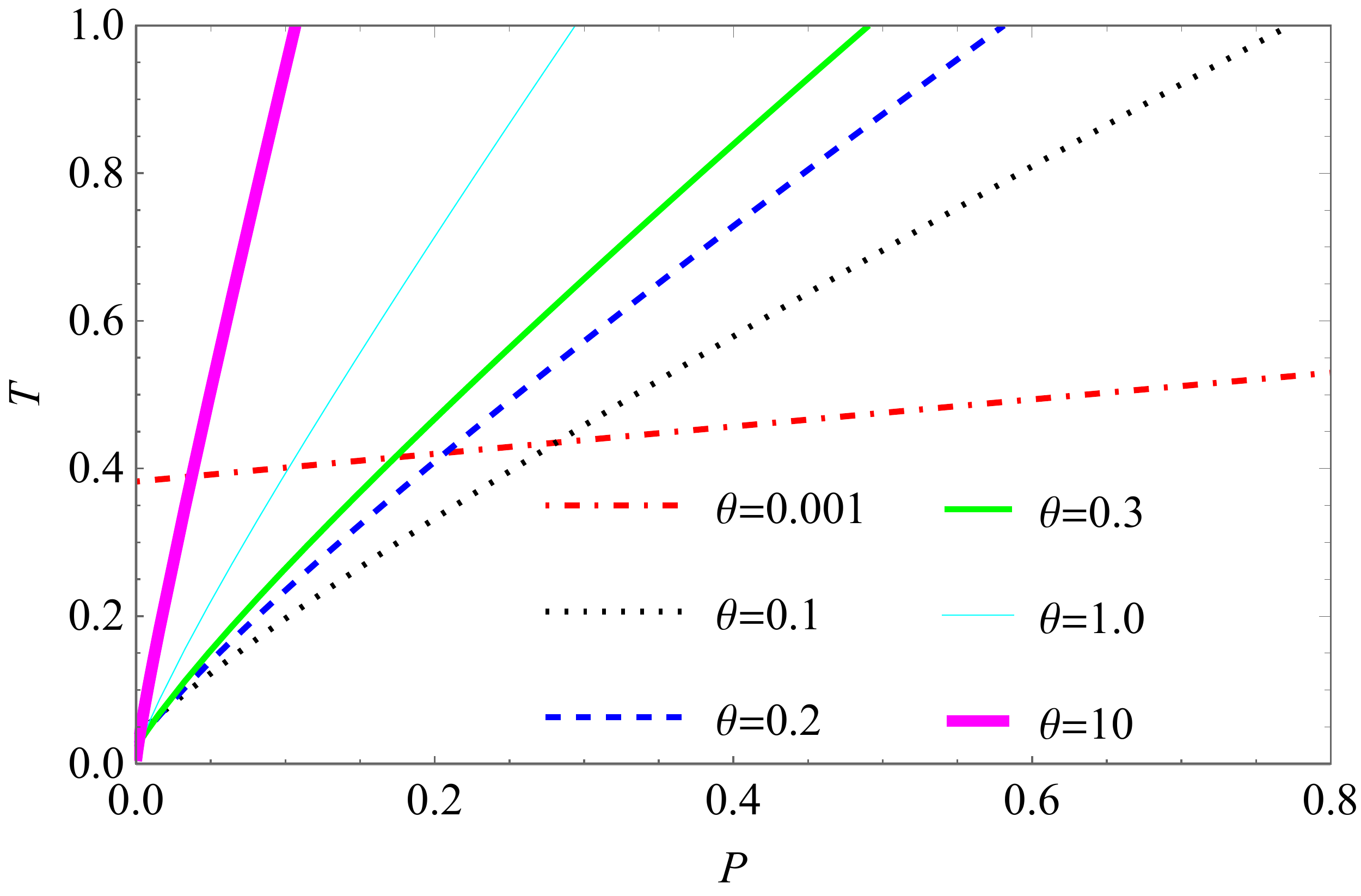} 
  \includegraphics[scale = 0.4]{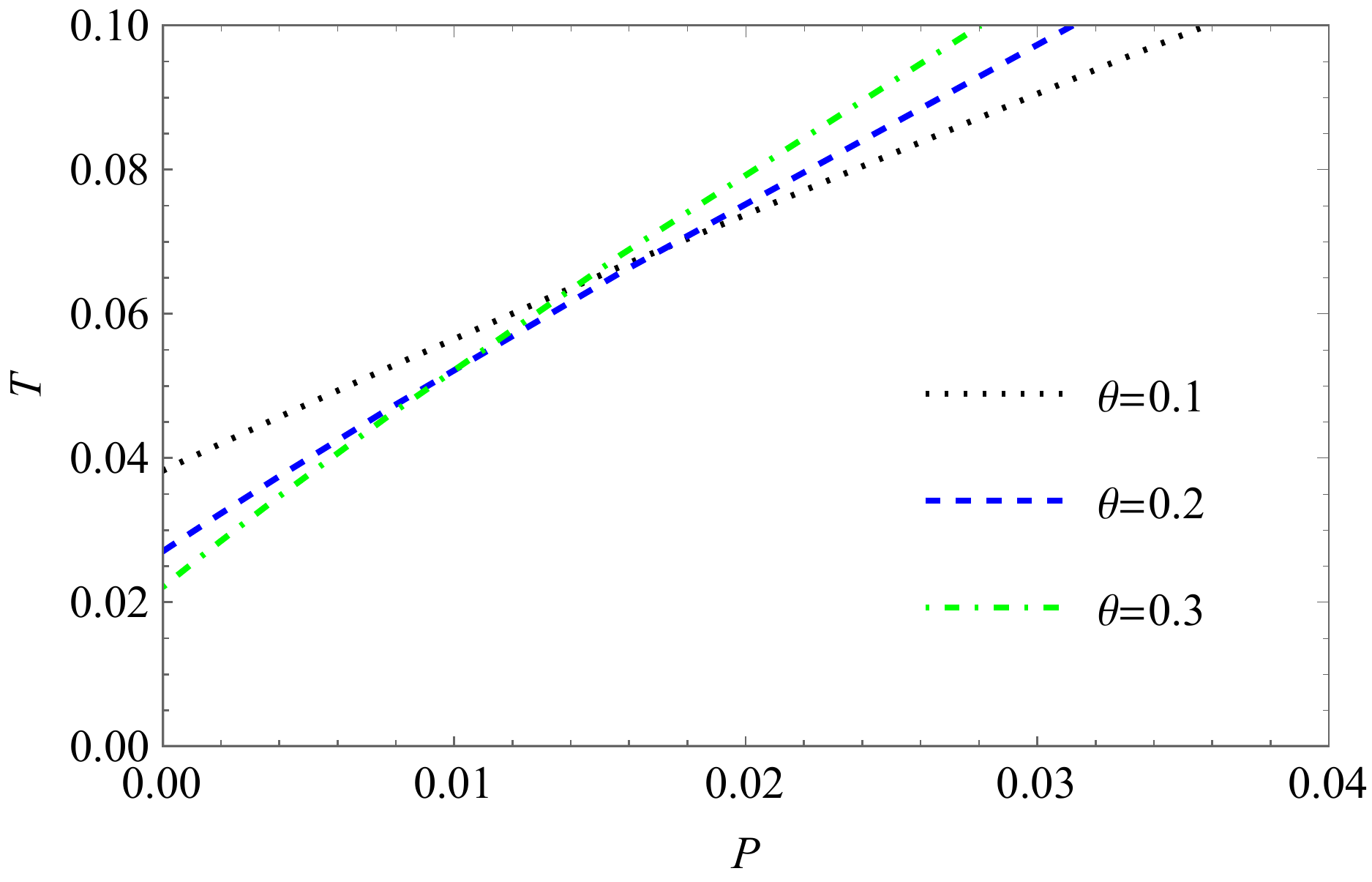} 
\caption{Inversion temperature curves for some values of the parameter $\theta$. In the left panel we plot a large  range of $P$, while in the right panel, we considered a small range for the scale for the pressure in order to capture the detail of the intercepts.}
\label{inversioncurves1}
\end{figure}

 To study the influence of the parameter $\theta$ on the inversion temperature against the pressure, we analyzed the range of $\theta$ from $10^{-3}$ to $10^{3}$. 
 In the left panel of Fig. \ref{inversioncurves1}, we plot $\theta$= 0.001, 0.1, 0.2, 0.3, 1 and 10 and see that increasing these values the slope of these curves also increase. This behavior also holds for higher values of $\theta$ and we checked it until $\theta $ reaches $10^3$. The shapes of the inversion temperature curves are similar to that of charged black holes in commutative theory.  

Notice that the limit $\theta \rightarrow 0$ should be taken with caution, since for $\theta = 0$, the system itself has no solution. 
In particular, for very small $\theta$ the slope is also very small, the intercept is high and the curve appears to become a straight line. 
 This seems  to be an indication that in the commutative limit $\theta\to 0$, there is no inversion temperature, as expected in AdS-Schwarzschild black holes. 

In the right panel of Fig. $\ref{inversioncurves1}$, one can see the behavior of the curves for small values of the inversion pressure. We are interested in finding the minimum inversion temperature ($P = 0$), which was one of the values investigated in the first paper on the subject \cite{Okcu:2016tgt}. As one can note, as the parameter $\theta$ increases, $T_i^{\text{min}}$ decreases. This is also the same behaviour for the charged black hole in general relativity.

\begin{figure}[!ht]
  \centering
  \includegraphics[scale = 0.40]{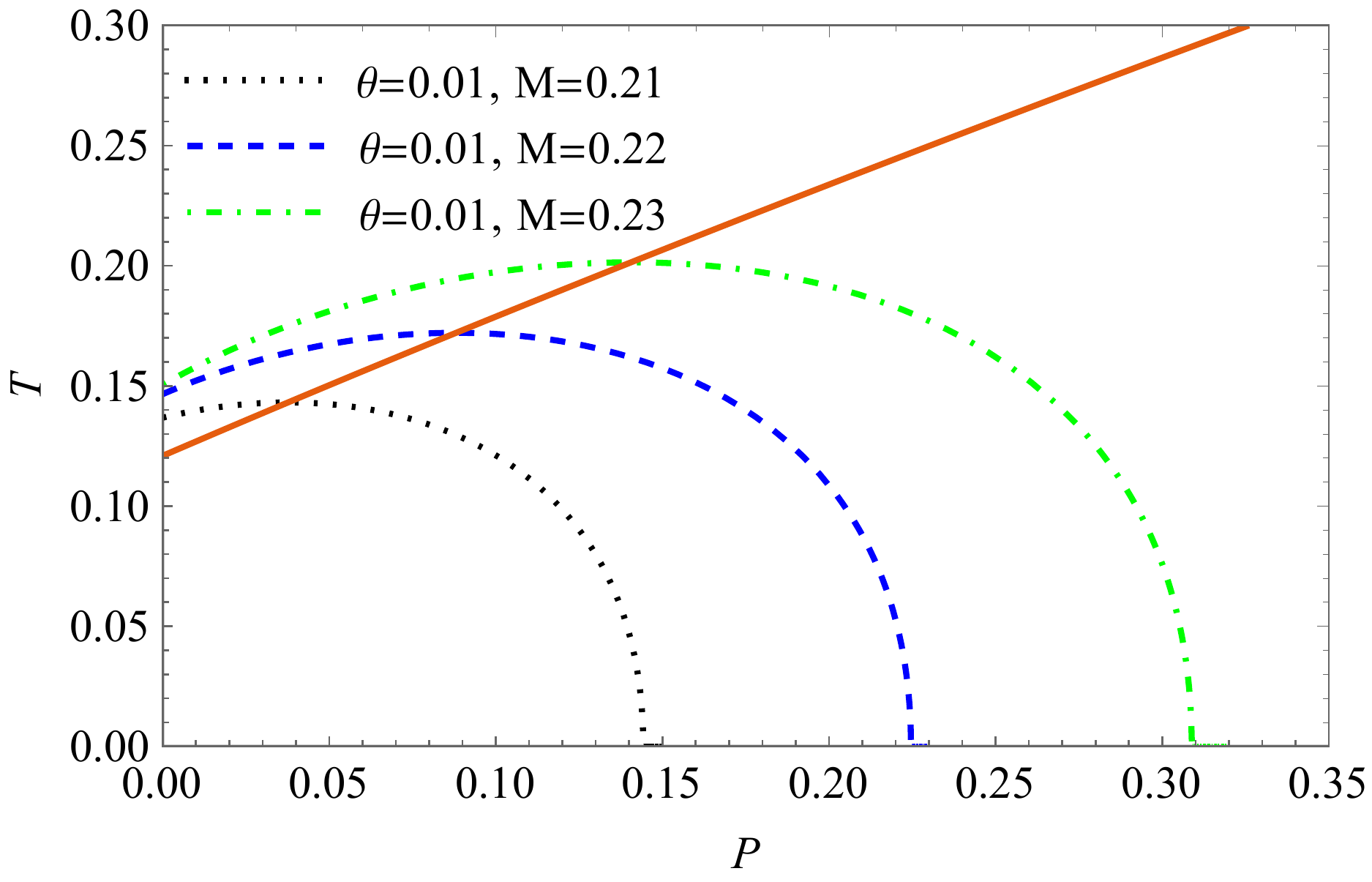} 
  \includegraphics[scale = 0.45]{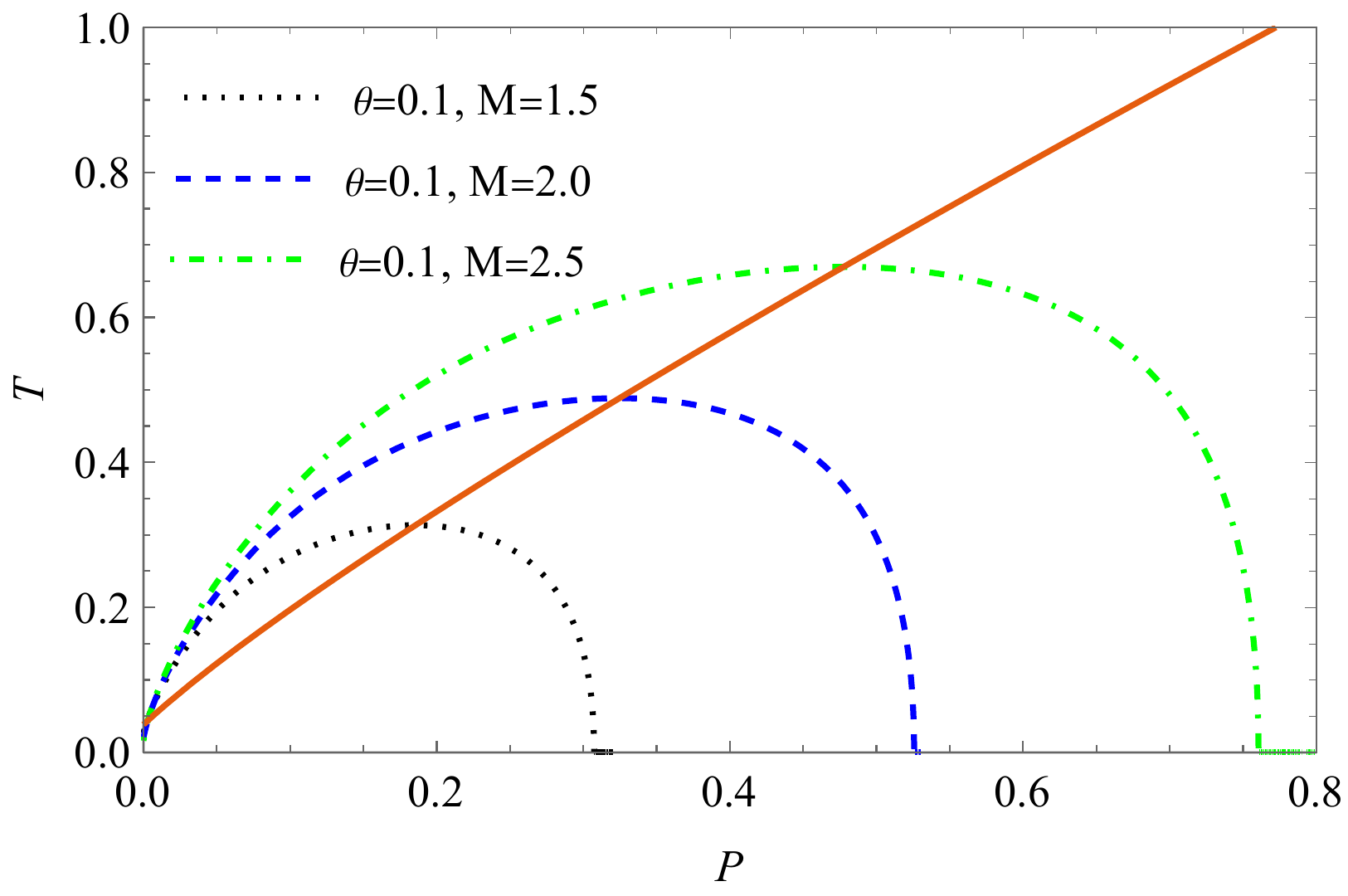} 
\caption{The dashed, dotted, and dot-dashed lines are the isenthalpics of the system, for different values of $M$ and the solid line is the inversion temperature curve. On the left, $\theta = 0.01$, and on the right, $\theta = 0.1$.}
\label{figIsenthalpics}
\end{figure}

We also plot some isenthalpics of the system, i.e., when the mass is constant. We made such a plot for different values for mass and the parameter $\theta$, as is show in figure \ref{figIsenthalpics}. The solid  line is the inversion temperature curve for the given parameter $\theta$, which intersects the isenthalpics at their maxima. These curves also resemble the corresponding ones for charged black holes. This feature still remains the same for higher values of $\theta$, but in this case one should also increase the values of the black hole mass.

\section{Conclusions}\label{conc}

In this paper, we studied the Joule-Thomson expansion for a noncommutative black hole. Such an expansion is a thermodynamic process where the temperature of the  system, usually a gas, varies with pressure, and may heat or cool. For real gases, there is usually an inversion temperature (and inversion pressure) where the gas stops heating and starts to cool, or vice versa. For black holes, this inversion only occurs in the presence of some type of charge, such as an electrical charge, for example. In the case of a Schwarzschild black hole, this reversal from heating to cooling does not occur.

We did this study by dealing with the question numerically, plotting the temperature inversion curves for the noncommutative black hole, for different values of the noncommutativity parameter. We also plot the isenthalpics of the system to capture the whole thermodynamic scenario.

Finally, our aim this work was to verify what happens when the black hole is uncharged but not commutative. We found that not only does inversion temperature occur, but the role of the noncommutativity parameter is similar to that of a charge, i.e., the higher its value, the lower the minimum temperature of the system for which the inversion occurs.



\begin{acknowledgments}
The  authors  would  like  to  thank  Danning  Li  for  numerical  support. J.P.M.G. is supported by Conselho Nacional de Desenvolvimento Científico e Tecnológico (CNPq) under Grant No. 151701/2020-2.
H.B.-F. is partially supported by Conselho Nacional de Desenvolvimento Científico e Tecnológico (CNPq) under Grant No. 311079/2019-9. This study was financed in part by the Coordenação de
Aperfeiçoamento de Pessoal de Nível Superior (CAPES), finance code 001. 
\end{acknowledgments}


\end{document}